\documentclass{article} % For LaTeX2e
\usepackage{iclr2027_conference,times}
\usepackage{graphicx}

\usepackage{amssymb} 
\usepackage{amsmath}
\usepackage{subcaption}
\usepackage{listings}
\usepackage{tikz}
\usetikzlibrary{arrows.meta,calc,positioning}
\usepackage{natbib}

\usepackage{amsmath,amsfonts,bm}

\def\eqref#1{equation~\ref{#1}}
\def\1{\bm{1}}

\DeclareMathAlphabet{\mathsfit}{\encodingdefault}{\sfdefault}{m}{sl}
\SetMathAlphabet{\mathsfit}{bold}{\encodingdefault}{\sfdefault}{bx}{n}

\usepackage{hyperref}
\usepackage{url}

\title{Learning and interpreting policies for\\simultaneous entanglement requests in\\quantum networks}
\definecolor{compcolor}{RGB}{128,190,110}
\definecolor{commcolor}{RGB}{70,135,205}
\definecolor{physicalcolor}{RGB}{76,114,176}
\definecolor{virtualcolor}{RGB}{190,55,85}
\definecolor{requiredcolor}{RGB}{105,75,160}

\definecolor{compcolor}{RGB}{128,190,110}
\definecolor{commcolor}{RGB}{70,135,205}
\definecolor{activecolor}{RGB}{55,105,180}
\definecolor{inactivecolor}{RGB}{155,155,155}
\definecolor{virtualcolor}{RGB}{190,55,85}
\definecolor{experimentcolor}{RGB}{105,75,160}
\definecolor{lockcolor}{RGB}{235,155,45}

\tikzset{
  logical qubit/.style={
    circle,
    draw=black,
    fill=compcolor!55,
    minimum size=6mm,
    inner sep=0pt,
    font=\small
  },
  processor/.style={
    rounded corners=2pt,
    draw=black!65,
    fill=black!3,
    minimum width=1.5cm,
    minimum height=1.25cm
  },
  computation qubit/.style={
    circle,
    draw=black,
    fill=compcolor!55,
    minimum size=6mm,
    inner sep=0pt,
    font=\small
  },
  communication qubit/.style={
    circle,
    draw=commcolor!90!black,
    fill=commcolor!30,
    minimum size=2.4mm,
    inner sep=0pt
  },
  required interaction/.style={
    draw=requiredcolor,
    line width=1.1pt
  },
  missing interaction/.style={
    draw=virtualcolor,
    line width=1.3pt
  },
  physical channel/.style={
    draw=physicalcolor,
    double=white,
    double distance=1.6pt,
    line width=0.7pt
  },
  swapping channel/.style={
    draw=commcolor!80!black,
    double=white,
    double distance=1.8pt,
    line width=1.2pt
  },
  virtual link/.style={
    draw=virtualcolor,
    dashed,
    line width=1.4pt
  }
}
\tikzset{
  graph/.style={
    x=4mm,
    y=4mm,
    line cap=round,
    line join=round
  },
  vertex/.style={
    circle,
    draw,
    minimum size=2.4mm, % radius = 1.2 mm = 0.12 cm
    inner sep=0pt,
    outer sep=0pt
  }
}

\tikzset{
  processor/.style={
    rounded corners=2pt,
    draw=black!65,
    fill=black!3,
    minimum width=1cm,
    minimum height=0.78cm
  },
  computation qubit/.style={
    circle,
    draw=black,
    fill=compcolor!60,
    minimum size=3.8mm,
    inner sep=0pt
  },
  communication qubit/.style={
    circle,
    draw=commcolor!90!black,
    fill=commcolor!30,
    minimum size=1.2mm,
    inner sep=0pt
  },
  network node/.style={
    circle,
    draw=black!70,
    fill=white,
    minimum size=6mm,
    inner sep=0pt,
    font=\scriptsize
  },
  experiment node/.style={
    circle,
    draw=experimentcolor,
    fill=experimentcolor!15,
    minimum size=5.5mm,
    inner sep=0pt,
    font=\scriptsize
  },
  active link/.style={
    draw=activecolor,
    line width=1.15pt
  },
  inactive link/.style={
    draw=inactivecolor,
    dashed,
    line width=0.9pt
  },
  virtual link/.style={
    draw=virtualcolor,
    dashed,
    line width=1.3pt
  },
  experiment edge/.style={
    draw=experimentcolor,
    line width=1.1pt
  },
  locked halo/.style={
    draw=lockcolor!45,
    line width=4pt
  },
  state box/.style={
    rounded corners=1.5pt,
    draw=black!50,
    fill=black!3,
    align=center,
    font=\tiny,
    inner sep=2.5pt
  }
}
\author{
\textbf{Leon Rode}\textsuperscript{1}
\qquad
\textbf{Sumeet Khatri}\textsuperscript{2}
\qquad
\textbf{Supartha Podder}\textsuperscript{1}
\\[0.5em]
\textsuperscript{1}Department of Computer Science,
Stony Brook University
\\
\textsuperscript{2}Department of Computer Science and\\~~Center for Quantum Information Science and Engineering,
Virginia Tech
\\[0.25em]
\texttt{\{lrode, supartha\}@cs.stonybrook.edu}
\qquad
\texttt{skhatri@vt.edu}
}

\iclrfinalcopy % Uncomment for camera-ready version, but NOT for submission.
\begin{document}

\maketitle

\begin{abstract}
Future quantum networks will make use of entanglement to perform numerous tasks, such as sending quantum information over long distances, distributed quantum computing, and quantum sensing. In general, these tasks will need to be performed simultaneously in various regions of a network, while minimizing resources and latency. We will thus require policies for scheduling link-level entanglement resources, and using the link-level entanglement to create various forms of multipartite entanglement required for every task. In this work, we address this problem using reinforcement learning. 
% We give a probabilistic model of a quantum network and formulate a problem in which a set of experiments, described by a topology, are to be executed on the network by performing entanglement swapping as a subroutine for generating necessary links absent in the host network. 
We formulate a Markov Decision Process for the problem and use double deep Q-networks (DQN) with Message Passing Neural Networks (MPNNs), experience replay buffers, and curriculum training to obtain policies. The key physical parameter is the probability of link-level entanglement generation, i.e., the link activation probability. We show that our policies maintain 100\% success for up to 71\% lower link activation probability than the baseline heuristics for a set of physically relevant network topologies. We then examine an additional constraint where experiment (task) placements are restricted to specific hardware types and demonstrate a similar advantage in performance over heuristics, with our policy maintaining at least an 80\% success rate for up to a 59\% lower link activation probability. Finally, to understand how our policies obtain their advantage, we explore methods to interpret the learned policy by defining metrics enabling conclusions to be drawn about the model's behavior and by tasking a large language model (LLM) to derive a novel heuristic given example actions taken by the DQN-trained policy. We find that the LLM heuristic performs similarly to the DQN-trained policy in performance, indicating a promising method for interpretable policy extraction for large quantum networks, where direct training becomes computationally expensive.
\end{abstract}

\section{Introduction}

One of the grand visions of quantum science and technologies is the realization of a quantum internet~\citep{WEH18}, a globally interconnected network of networked quantum computers that will, in conjunction with the existing (classical) internet, enable a host of applications, including those with known quantum advantages. Such applications include quantum computation, sensing, interferometry, and secure communication~\citep{networks_review,azuma2023repeatersRMP}.

While much work remains to realize this vision~\citep{awschalom2021interconnects}, networked quantum computation is an especially intriguing idea already, due to the fact that it could be used to scale up quantum computation itself. One of the current challenges hindering large-scale quantum computation on a single quantum computer is the qubit overhead of error correction. Currently, monolithic quantum computer architectures exist that can only support roughly hundreds of physical qubits, while qubit overhead for fault-tolerant error correction imposes a requirement of many more qubits~\citep{zhao2026ultrahighratequantumerrorcorrection}. Scaling beyond these values in monolithic architectures, while enabling fault tolerance via error correction, remains a challenge. Using entanglement to connect many such computers offers a path towards large-scale quantum computation. In fact, a successful demonstration of Grover's algorithm and the execution of other arbitrary two-qubit gates in a distributed setting has recently been shown~\citep{dqc}.

% A form of quantum internet that connects quantum memories of varying underlying platforms using shared entangled pairs of photons over long physical distances is becoming increasingly prevalent for distributed quantum computation \citep{networks_review}. 
% Distributed quantum computation offers a solution in which quantum computers, each supporting fewer qubits and thus requiring less error correction overhead, perform  part of the computation and communicate information among themselves through a quantum channel over which the entanglement is shared. Quantum gate teleportation offers a way to perform non-local operations over such a channel, providing a basis for distributed computation on physically distant set of quantum nodes. 

The global development of quantum network test beds is increasing~\citep{tokyo_photon,HKO+12,knaut2024entanglementnetwork,PHB+21,pompili2022demonstrationstack,chung2021illinois,chung2022illinois,nyc_photon}. In these networks, quantum nodes are physically connected using fiber optic cables that are later used to create entanglement pairs between neighboring nodes. However, not all quantum circuits can be directly executed on the network due to a mismatch between the quantum circuit pattern and the physical topology of the network. For example, if a network were constructed from four single-qubit quantum nodes connected in a ring-like fashion, it would not support a 4-qubit quantum circuit containing three two-qubit CNOT gates between $q_0$ and $q_1$, $q_1$ and $q_2$, and $q_2$ and $q_0$. In this case, a ``virtual'' link is needed to support one of the CNOT operations. To obtain a virtual link between distant nodes $u$ and $w$, both independently sharing entanglement with $v$, one performs entanglement swapping~\cite{ZZH93} between both pairs $u$-$v$ and $v$-$w$, resulting in a shared entanglement between $u$ and $w$. This process can be repeated to enable virtual links over greater distances. 

Previous research has examined linear programming \citep{khatri2022networkMDP,inesta2023optimal,fan2024optimized}, decision transformers, \citep{seok26endtoend}, and learning approaches \cite{ RL22,Le_2022,Haldar_2024,li2025optimising} to address the problem of efficiently resolving a queue of virtual link generation requests on the network in minimal time, while maximizing the fidelity of the resulting virtual link. However, in our work, we examine the problem of fulfilling not the queue of virtual link requests but a given queue of ``jobs'' or ``experiments'' that end-users may request to execute. Each such experiment is a particular \textit{subgraph} of physical and/or virtual links in the network, which could correspond to a multipartite state used for quantum sensing or for a distributed quantum computation. Given a set of such subgraphs, each requiring a given duration of time to execute, our goal is to  determine an optimal policy that generates all subgraphs in the network, minimizing the time required. This formulation was examined in \cite{ni2026advanced}, in which they propose heuristics to efficiently schedule the experiments. For this problem, virtual link generation is a \textit{subroutine} necessary to \textit{enable} the execution of an arbitrary distributed quantum circuit. In the quantum analogy of the classical network stack, fulfilling virtual link generation requests is contained in the so-called connectivity layer, whereas the problem we examine lies in the link layer as it involves requesting the required virtual links from the connectivity layer to schedule the experiments \citep{network_stack}. The problem has real-world implications in that broader usage of a mature quantum internet will result in a high number of end-users seeking to execute circuits of inherently variable topology demands on a network limited to a static physical topology. 
 
We also examine an additional constraint in which a quantum circuit is confined to specific quantum nodes in the network for reasons such as hardware compatibility. The network operator will need to efficiently schedule the execution of the circuits on the network while maximizing fidelity over them and satisfying the constraint. 

%\SK{Here we should have a paragraph summarizing the main results.}

%\LR{We can remove this paragraph in favor of main results paragraph}
We find that our models outperform baselines in terms of robustness to changes in link activation probability, measured by the success rate and mean episode length. In particular, we find that our models succeed more frequently and more efficiently in more lossy environments than the baselines, indicating that lossy, near-term quantum network implementations can benefit from DQN for multipartite entanglement scheduling tasks. Additionally, this performance extends to a realistic constraint in which certain nodes in the network may be preferred by the network's users, which limits their experiment's possible locations on the network, ultimately increasing the role bottleneck nodes might have. Lastly, by prompting an LLM to infer a policy from example trajectories generated by the DQN model, we were able to extract a policy--expressible using natural language--that performs nearly as well as our DQN model, showing the significance of LLMs in extracting performant yet interpretable policies given a DQN training process.

%In Section \ref{networkmodel}, we describe our model of the quantum network that includes probabilistic physical link activation, the virtual link generation subroutine, and the specific conditions for quantum circuit placement on the network. We then formulate the environment as a Markov Decision Process. In Section \ref{rl_framework}, we describe our implementation of double deep Q-networks combined with experience replay buffers and curriculum training to derive a more noise-robust policy compared to baseline heuristics described in Section \ref{setup}. In Section \ref{results}, we show superior performance of our learned policy compared to the baseline heuristics across three different network topologies, with and without the additional constraint of hardware compatibility. Lastly, in Section \ref{behavior_profiling}, we profile the behavior of the learned policy and baseline heuristics, aiding in extracting the decision rules of the learned policy. We conclude with an  examination of using large language models (LLMs) to derive a novel, and more importantly, explainable heuristic for solving the task, and show that a frontier LLM was capable of devising a heuristic that shows similar robustness compared to the Q-network-based policy, paving the way for further LLM usage in model explainability.

\section{Network Model and Problem Formulation}\label{networkmodel}

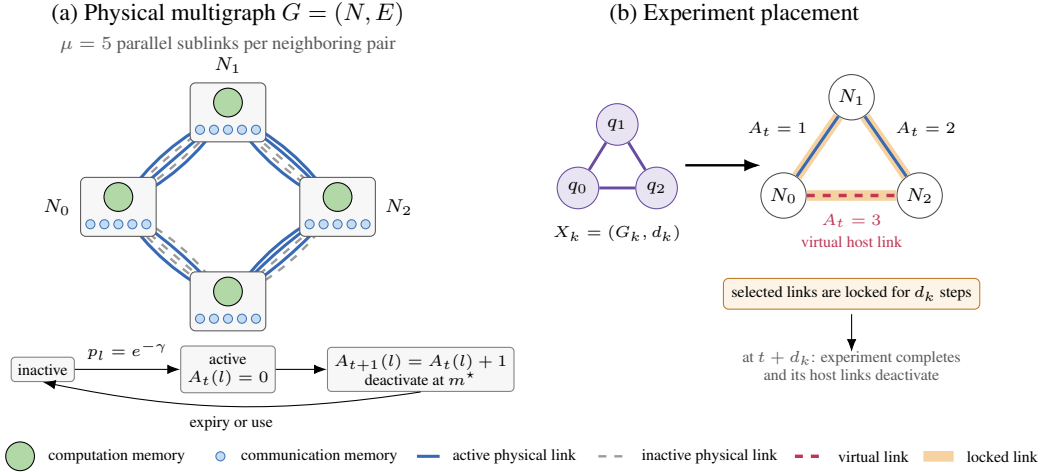
\begin{figure*}[t]
\centering
\begin{tikzpicture}[font=\small]

% ============================================================
% (a) Physical multigraph and link dynamics
% ============================================================
\begin{scope}[shift={(0,0)}]

  \node[align=center] at (0,2.55)
    {(a) Physical multigraph $G=(N,E)$};

  \node[font=\scriptsize, text=black!65] at (0,2.15)
    {$\mu=5$ parallel sublinks per neighboring pair};

  \coordinate (p0) at (-1.45, 0);
  \coordinate (p1) at ( 0.00, 1.25);
  \coordinate (p2) at ( 1.45, 0);
  \coordinate (p3) at ( 0.00,-1.25);
    
  % N0 -- N1
\foreach \linkstyle/\angle in {
  active link/14,
  inactive link/7,
  active link/0,
  inactive link/-7,
  active link/-14
}{
  \draw[\linkstyle]
    (p0) to[bend left=\angle] (p1);
}

% N1 -- N2
\foreach \linkstyle/\angle in {
  active link/14,
  active link/7,
  inactive link/0,
  inactive link/-7,
  active link/-14
}{
  \draw[\linkstyle]
    (p1) to[bend left=\angle] (p2);
}

% N2 -- N3
\foreach \linkstyle/\angle in {
  inactive link/14,
  active link/7,
  inactive link/0,
  active link/-7,
  active link/-14
}{
  \draw[\linkstyle]
    (p2) to[bend left=\angle] (p3);
}

% N3 -- N0
\foreach \linkstyle/\angle in {
  active link/14,
  inactive link/7,
  inactive link/0,
  active link/-7,
  inactive link/-14
}{
  \draw[\linkstyle]
    (p3) to[bend left=\angle] (p0);
}

  % Quantum-node containers
  \node[processor] (N0) at (p0) {};
  \node[processor] (N1) at (p1) {};
  \node[processor] (N2) at (p2) {};
  \node[processor] (N3) at (p3) {};

  % Computation and communication memories
  \foreach \n in {N0,N1,N2,N3}{
    \node[computation qubit]
      at ($(\n.center)+(0,0.13)$) {};

    %\node[communication qubit]
    %  at ($(\n.center)+(-0.23,-0.23)$) {};

    %\node[communication qubit]
    %  at ($(\n.center)+(0.23,-0.23)$) {};
    \foreach \x in {-0.37,-0.185,0,0.185,0.37}{
  \node[communication qubit]
    at ($(\n.center)+(\x,-0.23)$) {};
}

  }

  % Network-node labels
  \node[font=\scriptsize\bfseries, left=0mm of N0]  {$N_0$};
  \node[font=\scriptsize\bfseries, above=0.0mm of N1] {$N_1$};
  \node[font=\scriptsize\bfseries, right=0mm of N2] {$N_2$};
  \node[font=\scriptsize\bfseries, below=1mm of N3] {$N_3$};

  % Link-state evolution
  \node[state box] (inactive) at (-2.45,-2.15)
    {inactive};

  \node[state box] (active) at (0,-2.15)
    {active\\$A_t(l)=0$};

  \node[state box] (aging) at (2.55,-2.15)
    {$A_{t+1}(l)=A_t(l)+1$\\deactivate at $m^\star$};

  \draw[-{Latex[length=1.7mm]}, thin]
    (inactive) --
    node[above, font=\tiny] {$p_l=e^{-\gamma}$}
    (active);

  \draw[-{Latex[length=1.7mm]}, thin]
    (active) -- (aging);

  \draw[-{Latex[length=1.7mm]}, thin]
    (aging.south) to[bend left=10]
    node[below, font=\tiny] {expiry or use}
    (inactive.south);
\end{scope}

% ============================================================
% (b) Experiment placement and locking
% ============================================================
\begin{scope}[shift={(6.7,0)}]

  \node[align=center] at (0,2.55)
    {(b) Experiment placement};

  % Experiment topology G_k
  \begin{scope}[shift={(-1.55,0.65)}, scale=0.8]
    \node[experiment node] (q0) at (-0.65,-0.5) {$q_0$};
    \node[experiment node] (q1) at ( 0.00,0.55) {$q_1$};
    \node[experiment node] (q2) at ( 0.65,-0.5) {$q_2$};

    \draw[experiment edge] (q0) -- (q1);
    \draw[experiment edge] (q1) -- (q2);
    \draw[experiment edge] (q0) -- (q2);

    \node[font=\tiny, align=center] at (0,-1.25)
      {$X_k=(G_k,d_k)$};
  \end{scope}

  % Injective experiment placement
  \draw[-{Latex[length=2.2mm]}, line width=0.9pt]
    (-0.65,0.55) -- (0.35,0.55);

  % Relevant portion of the active host graph
  \node[network node] (H0) at (0.65,0.15) {$N_0$};
  \node[network node] (H1) at (1.55,1.45) {$N_1$};
  \node[network node] (H2) at (2.45,0.15) {$N_2$};

  % Orange halos indicate links locked by the experiment
  \draw[locked halo] (H0) -- (H1);
  \draw[locked halo] (H1) -- (H2);
  \draw[locked halo] (H0) -- (H2);

  % Physical and virtual host links
  \draw[active link]
    (H0) -- node[midway, above left, font=\tiny]
    {$A_t=1$} (H1);

  \draw[active link]
    (H1) -- node[midway, above right, font=\tiny]
    {$A_t=2$} (H2);

  \draw[virtual link]
    (H0) -- node[
      midway,
      below=1mm of H0,
      font=\tiny,
      text=virtualcolor
    ] {$A_t=3$} (H2);

  \node[
    font=\tiny,
    text=virtualcolor,
    below=1mm of H0,
    xshift=9mm
  ] {virtual host link};

  \node[
    rounded corners=2pt,
    draw=lockcolor!80!black,
    fill=lockcolor!12,
    align=center,
    font=\tiny,
    inner sep=3pt
  ] at (1.55,-1.15)
    {selected links are locked for $d_k$ steps};

  \draw[-{Latex[length=1.7mm]}, thin]
    (1.55,-1.42) -- (1.55,-1.92);

  \node[
    align=center,
    font=\tiny,
    text=black!65
  ] at (1.55,-2.1)
    {at $t+d_k$: experiment completes\\
     and its host links deactivate};
\end{scope}

% ============================================================
% Legend
% ============================================================
\begin{scope}[shift={(-1.25,-3.3)}]
  \node[computation qubit] at (-1.5,0) {};
  \node[anchor=west, font=\tiny] at (-1.25,0)
    {computation memory};

  \node[communication qubit] at (1.15,0) {};
  \node[anchor=west, font=\tiny] at (1.3,0)
    {communication memory};

  \draw[active link] (3.75,0) -- (4.05,0);
  \node[anchor=west, font=\tiny] at (4.10,0)
    {active physical link};

  \draw[inactive link] (6.15,0) -- (6.55,0);
  \node[anchor=west, font=\tiny] at (6.60,0)
    {inactive physical link};

  \draw[virtual link] (8.75,0) -- (9.15,0);
  \node[anchor=west, font=\tiny] at (9.20,0)
    {virtual link};

  \draw[locked halo] (10.45,0) -- (10.85,0);
  \node[anchor=west, font=\tiny] at (10.90,0)
    {locked link};
\end{scope}

\end{tikzpicture}

\caption{
(a) A quantum network composed of $N=4$ quantum memories organized as a ring, with $\mu=5$ parallel sublinks between neighboring nodes. Links move from an inactive state to an active state with probability $p_l = e^{-\gamma}$ after which their age increments from zero to $m^\star$. (b) Entanglement swapping is performed across link $N_0$-$N_1$ and link $N_1$-$N_2$ with ages 1 and 2 respectively, establishing a virtual link between $N_0$-$N_2$ with age 3. This enables experiment $X_k$ to be placed on host nodes $N_0, N_1, N_2$. The links are consumed for $d_k$ steps before being released.}
\label{fig:virtual-link-placement}
\end{figure*}

Let $G$ be a network of nodes $N$ and a set of physical links, $E$, representing optical connections between nodes. Let there be $\mu$ parallel links (sublinks) between neighboring nodes, enabled by multiple quantum memories operating on each node. Thus $G$ is a multi-graph. Hereafter, a physical link between two nodes refers to any of the $\mu$ sublinks between them. 

The network evolves in discrete time steps $t \in \mathbb{N}_0$. $E$ is partitioned into active and inactive physical links. At each time step, an inactive physical link attempts to activate, succeeding with probability $p_l = e^{-\gamma}$, where $\gamma$ represents the attenuation coefficient of the link. Upon successful activation, the link $l$ initializes at age 0 and increments each step so that its age satisfies $
A_t(l) \in \{0, 1, ..., m^\star -1 \}$,
remaining active for at most $m^\star \in \mathbb{N}$ time steps before deactivating.
 
A ``virtual'' link may be generated between two non-neighboring nodes via entanglement swapping over their shortest connecting path of links (both physical and virtual), $P$. Given $P$, if $
\sum_{l \in P} A_t(l) < m^\star$, a virtual link may be generated between the path's endpoints, whose age is $\sum_{l\in P} A_t(l)$. Once generated, all $l \in P$ become inactive. Hence, in addition to $E$, we have $V = (N \times N) \backslash E$ of generable virtual links in the network $G$. We assume that entanglement swapping succeeds deterministically.

Let $\mathcal{X}$ represent the set of experiments to be executed across distributed nodes. Each experiment defines a topology of shared entangled pairs between quantum nodes. Gate speeds, teleportation overhead, classical communications, and other factors constitute the duration of the experiment. These parameters define an individual experiment $X_k \in \mathcal{X}$, $X_k = (G_k, d_k)$, where $G_k = (N_k, E_k)$ is the required topology of experiment $k$ and $d_k>0$ is its duration. Figure~\ref{fig:virtual-link-placement} describes the network dynamics on a four node graph, including virtual link generation and experiment placement.

Let $\tilde{G}_t = (N, \tilde{E}_t \cup V_t)$ be named the ``active'' graph, whose set of edges is composed of active physical links $\tilde{E}_t$ and generated virtual links $V_t$ at time $t$. An experiment can be ``placed'' onto $\tilde{E}_t \cup V_t$ if (i) $G_k$ is subgraph isomorphic to $\tilde{G}_t$ (that is, there exists an injective mapping $f:N_k \rightarrow N$ such that $(u, v) \in E_k \Rightarrow (f(u), f(v)) \in \tilde{E}_t \cup V_t$), and (ii) for all active host links $l \in \tilde{E}_t \cup V_t$, $A_t(l) \leq m^\star - d_k$, i.e. no link will deactivate before $d_k$ time steps pass.
Once $X_k$ is placed, all host links are considered locked for the duration $d_k$ and cannot be re-assigned or used for virtual link generation. Upon completion of $X_k$ at step $t+d_k$, the host links are deactivated, after which they repeatedly try to activate with success probability $p_l$. 

We also consider the constraint in which an experiment $X_k$ may limit its placement to a subset of host nodes. As the quantum network may be heterogeneously composed of various types of quantum memory, a user may desire to delay the placement of their experiment until the memory with desired characteristics can be utilized. Define $c(u)$ as the set of ``colors'' of node $u$. Each host node $u \in N$ maintains a non-empty set $c(u)$, while the color set of an experiment node $v \in N_k$, $c(v)$, may be empty, indicating hardware agnosticism. The placement of an experiment is only valid under the same conditions as previously described, with the addition that the mapped nodes $v \rightarrow f(v)$ satisfy $c(v) \cap c(f(v)) \neq \emptyset \lor c(v) = \emptyset$, i.e. the colors of the host and experiment nodes agree so long as an experiment node has color preferences.

\subsection{Markov Decision Process Formulation}

We use reinforcement learning to train a model to minimize the time to place all experiments in $\mathcal{X}$. The underlying mathematical model, the Markov Decision Process (MDP), describes the state of the environment at time $t$, and the set of valid actions. The agent receives at each time step: (1) the number of memories available in each node normalized by $\mu$; (2) the active sublinks in the topology and their age normalized by $m^\star$; (3) the time remaining for sublinks locked by an experiment; and (4) which experiments have been completed. The set of actions available to the agent is either to place an experiment, to generate a virtual link, or to wait (take no action). The validity of placements and virtual link generations is signaled using a binary action mask. After taking a specific action, the environment returns a reward equal to the sum of the relevant reward terms.

\paragraph{Subgraph edit distance-aware reward.} Placing an experiment $X_k = (G_k, d_k)$ at time $t$ requires $G_k$ to be subgraph isomorphic to $\tilde{G}_t$. We define a custom notion of \textit{subgraph edit distance} (SED) between $G_k$ and $\tilde{G}_t$ as the minimum number of additional edges needed in $G_k$ for it to be subgraph isomorphic to $\tilde{G}_t$. Assume that the model chooses to generate a virtual link at time $t$. Let $\Delta \operatorname{SED} = \operatorname{SED}_{t} - \operatorname{SED}_{t-1}$ be negative if the minimum SED between the topology of an unassigned experiment and any of its placements in $\tilde{G}_t$ decreased, where the minimum is calculated over all experiments and all placements. Then the reward received by the agent for virtual link generation, $r_{\text{vl}}$, is defined as
\begin{equation}
r_{\text{vl}} = \begin{cases}
    -r_{\text{pen}} & \text{if } \Delta \operatorname{SED} > 0 \\
    \begin{aligned}[t]
        &r_{\text{base}} - \alpha \operatorname{SED}_{t} - \beta(\Delta \operatorname{SED} + 1)
    \end{aligned} & \text{otherwise}
\end{cases}
\end{equation}
where $r_{\text{pen}}, r_{\text{base}}>0$ and $\alpha, \beta>1$ are tunable parameters. If the virtual link increases the minimum SED between any unassigned experiment topology and $\tilde{G}_t$, the agent is penalized by $r_{\text{pen}}$, representing an unhelpful generation. If the virtual link decreases the overall minimum SED, then the agent receives a reward composed of a base amount, a linear penalty proportional to the remaining distance ($\operatorname{SED}_t$), and a shaping reward for the magnitude of step progress ($\Delta \operatorname{SED}$). We clarify this choice through the following example. Consider two virtual link generations, both of which cause $\Delta \operatorname{SED} = -1$. Let Action $A$ reduce the overall SED from $4 \to 3$, while Action $B$ reduces it from $1 \to 0$. Under Action $A$, placing any experiment requires at least 3 additional virtual links. In contrast, Action $B$ enables the model to place the experiment in the next time step (assuming that the links did not expire). The weighting of $r_{\text{vl}}$ by $\operatorname{SED}_t$ incentivizes the model to prioritize virtual links that allow a quicker placement of the experiment, since the goal is to minimize the needed time.

\paragraph{Betweenness centrality-aware reward.} We introduce an additional reward depending on the maximum betweenness centrality of nodes along the path whose links were consumed to generate a virtual link. The betweenness centrality of a node $u$ measures the number of shortest paths, taken between all node pairs in the graph, that include $u$ along the path. A greater value indicates that a node acts more like a bottleneck. We compute its value based on the active edge set $\tilde{E}_t \cup V_t$ and normalize it against the maximum possible betweenness centrality out of all generable virtual links at that time step. If at time $t$, $V^+_t \subseteq V$ is the set of generable virtual links, $l \in V^+_t$ is the virtual link chosen, and $P(l)$ is the shortest path among $\tilde{E}_t \cup V_t$ between the endpoints of $l$, then
\begin{equation}
    r_\text{bottleneck} = r^\text{bottleneck}_\text{base} \times \frac{\max_{p \in P(l)} g(p)}{\max_{v \in V^+} \left\{\max_{q \in P(v)} g(q)\right\}} \in [0, r_\text{base}^\text{bottleneck}]
\end{equation}
is the reward given to the agent, where $g$ computes the betweenness centrality of a node. Here, $r_\text{base}^{\text{bottleneck}}>0$ is a hyperparameter.

\paragraph{Experiment complexity-aware reward for placement} When the agent chooses to place an experiment $X_k$ with edge set $E_k$, it receives a reward $r_{\text{exp}} = r_{\text{base}}^\text{exp}  \times \frac{|E_{k}|}{\kappa}$, where $|E_{k}|$ is the number of edges in $G_k$, $\kappa>0$ is a scaling parameter weighting the edge count, and $r_{\text{base}}^\text{exp}$ is an overall weighting parameter. Since more complex experiments require more intermediate virtual links, and thus more time to place, the greater penalty received in placing the experiment should be offset by a greater reward. A static reward would incentivize the agent to greedily place simpler experiments instead of normalizing against the required spatial resources.  

\paragraph{Episode length penalty}To incentivize the agent to minimize the overall time required to place all experiments in $\mathcal{X}$, we penalize the agent for each time step used with
$r_{\text{step}} <0$. 

\section{Reinforcement Learning Framework}\label{rl_framework}
We use double deep Q-learning to train our model. In this method, an online network (student) interacts with the environment through actions that a target network (teacher) evaluates, with the disagreement between the two models being the objective function to minimize.  Higher values of $\gamma$ increase the difficulty in placing experiments due to fewer simultaneous active links in a given interval of time. Therefore, training solely under high environment noise leads to difficulties in converging to an optimal policy, as any learning signal is subdued by the environment's noise. To improve policy performance at higher noise levels, we employ curriculum training with a replay buffer. Using curriculum training, the model learns a policy in low-noise conditions, which is then modified as it further trains in high-noise conditions. By exposing the model to low-noise environments, it is able to converge on a high-fidelity policy early on, while adjusting to simultaneously increasing stochasticity. We curriculum-train the model in 11 phases, each phase $\mathcal{P}$ corresponding to a specific $\gamma$ of 11 linearly-interpolated values of the interval $[1.5, 5.8]$. ($p_l$ decreases from $\approx 0.22$ to $\approx 0.003$ over the course of the curriculum.) If $\mathcal{P} > 1$, the replay buffer is seeded with episode roll-outs from the model from $\mathcal{P} - 1$. Then the online network is trained until a maximum step count is reached or phase mastery is achieved. Phase mastery is achieved after sufficient convergence and the demonstration of a mean success rate of 100\% over a window of 100 episodes. Finally, the saved weights are used to both warm-start and seed the next phase $\mathcal{P} + 1$.

We maintain a state transition buffer (containing tuples of the state observed, the action taken, and the reward received) with capacity for 100,000 transitions. The transitions from 500 episodes rolled-out by the online network are pushed to the buffer. These constitute an ``expert'' section of the buffer and are kept unmodified for the phase. Transitions generated using an $\epsilon$-greedy approach by the online network during its training are pushed to the buffer. During the update phase, a mini-batch of 256 samples is drawn from the buffer with the ratio of 25\% expert transitions to 75\% online network transitions, which is used to compute the gradient of the loss function and for updating the target network using Polyak updates as described in Appendix \ref{appendix:DDQN}. Since the model learns from graph-structured information, we employ a Message Passing Neural Network (MPNN) for topology-aware decision-making. Details of the MPNN are left for Appendix \ref{appendix:MPNN}.

\begin{figure*}[t]
\begin{subfigure}[c]{0.3\textwidth}
\centering
\begin{tikzpicture}[
  graph,
  baseline=(current bounding box.center)
]
  \node[vertex] (a) at (-0.707, 0) {};
  \node[vertex] (b) at (-0.707,-1) {};
  \node[vertex] (c) at (-1.707, 0) {};

  \node[vertex] (d) at ( 0.707, 0) {};
  \node[vertex] (e) at ( 1.707, 0) {};
  \node[vertex] (f) at ( 0.707,-1) {};

  \node[vertex] (g) at ( 0,     1) {};
  \node[vertex] (h) at (-0.707, 2) {};
  \node[vertex] (i) at ( 0.707, 2) {};

  \draw
    (a) -- (b)
    (a) -- (c)
    (a) -- (d)
    (a) -- (g)
    (d) -- (e)
    (d) -- (f)
    (d) -- (g)
    (g) -- (h)
    (g) -- (i);
\end{tikzpicture}
\end{subfigure}%
\hfill
%
% Second graph
\begin{subfigure}[c]{0.3\textwidth}
\centering
\begin{tikzpicture}[
  graph,
  baseline=(current bounding box.center)
]
  % Central path
  \node[vertex] (center) at ( 0, 0) {};
  \node[vertex] (left)   at (-1, 0) {};
  \node[vertex] (right)  at ( 1, 0) {};

  % Left diamond
  \node[vertex] (lefttop)    at (-2, 1) {};
  \node[vertex] (leftmiddle) at (-3, 0) {};
  \node[vertex] (leftbottom) at (-2,-1) {};

  % Right diamond
  \node[vertex] (righttop)    at (2, 1) {};
  \node[vertex] (rightmiddle) at (3, 0) {};
  \node[vertex] (rightbottom) at (2,-1) {};

  \draw
    (leftmiddle) -- (lefttop)
    (leftmiddle) -- (leftbottom)
    (lefttop) -- (left)
    (leftbottom) -- (left)
    (left) -- (center)
    (center) -- (right)
    (right) -- (righttop)
    (right) -- (rightbottom)
    (righttop) -- (rightmiddle)
    (rightbottom) -- (rightmiddle);
\end{tikzpicture}
\end{subfigure}%
\hfill
%
% Third graph
\begin{subfigure}[c]{0.3\textwidth}
\centering
\begin{tikzpicture}[
  graph,
  baseline=(current bounding box.center)
]
  % Bottom row
  \node[vertex] (a) at (-1,0) {};
  \node[vertex] (b) at ( 0,0) {};
  \node[vertex] (c) at ( 1,0) {};

  % Middle row
  \node[vertex] (d) at (-1,1) {};
  \node[vertex] (e) at ( 0,1) {};
  \node[vertex] (f) at ( 1,1) {};

  % Top row
  \node[vertex] (g) at (-1,2) {};
  \node[vertex] (h) at ( 0,2) {};
  \node[vertex] (i) at ( 1,2) {};

  \draw
    % Horizontal edges
    (a) -- (b) -- (c)
    (d) -- (e) -- (f)
    (g) -- (h) -- (i)

    % Vertical edges
    (a) -- (d) -- (g)
    (b) -- (e) -- (h)
    (c) -- (f) -- (i);
\end{tikzpicture}
\end{subfigure}
    \caption{The starlink (left), dumbbell (center), and grid (right) topologies.}
    \label{fig:topologies}
\end{figure*}
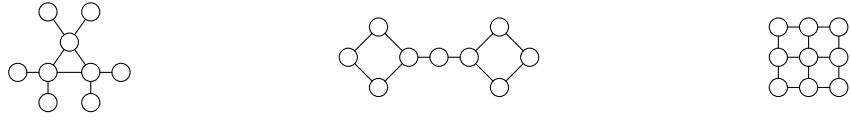

\section{Experimental Setup}\label{setup}

\subsection{Evaluation Topologies}

We train three models to operate in three topologies as shown in Figure \ref{fig:topologies}. The starlink topology shown represents a hub of three independent networks linked together. The dumbbell topology simulates two networks bridged by a bottleneck node. The grid topology is more dense, representing a network that may be implemented in urban settings or in a data center. In all evaluation settings, we set $|N| = 9$, $m^\star = 52$, $\mu=5$, and $\mathcal{X} = \{(K_4, 1), (K_4, 1)\}$,
where $K_4$ is the complete graph of 4 nodes. We truncate an episode if either experiment remains unplaced after 200 time steps. We consider the episode to be successful if it is not truncated.
\subsection{Baseline Heuristic Algorithms}

To benchmark the performance of our models, we compare their performance against baseline heuristics. The first baseline algorithm is a greedy approach we call \textsc{AgeCriticalFirst}, which places a valid experiment whose corresponding host links expire the soonest, followed by generating the virtual link with the least resulting age.  The second baseline algorithm is another greedy approach we call \textsc{ShortestHopFirst}, which places the first experiment in the action index, followed by generating the virtual link whose physical path distances is shortest. Whereas \textsc{AgeCriticalFirst} operates greedily with respect to time, \textsc{ShortestHopFirst} operates greedily with respect to space. The third heuristic, \textsc{DegreeCentricThresholdRouting} (\textsc{DCTR}), places the first experiment in the action index, followed by generating the virtual link whose centrality of its endpoints $u, v \in N$, $S = \deg(u) + \deg(v)$ is greatest.

\subsection{Color Constraint}\label{color_constraint}

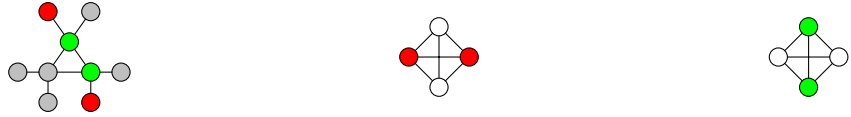
\begin{figure*}[h!]
\centering

% Left graph
\begin{subfigure}{0.3\linewidth}
\centering
\begin{tikzpicture}[
  graph,
  baseline=(current bounding box.center)
]
  \node[vertex, fill=lightgray] (a) at (-0.707, 0) {};
  \node[vertex, fill=lightgray] (b) at (-0.707,-1) {};
  \node[vertex, fill=lightgray] (c) at (-1.707, 0) {};

  \node[vertex, fill=green]     (d) at ( 0.707, 0) {};
  \node[vertex, fill=lightgray] (e) at ( 1.707, 0) {};
  \node[vertex, fill=red]       (f) at ( 0.707,-1) {};

  \node[vertex, fill=green]     (g) at ( 0,     1) {};
  \node[vertex, fill=red]       (h) at (-0.707, 2) {};
  \node[vertex, fill=lightgray] (i) at ( 0.707, 2) {};

  \draw
    (a) -- (b)
    (a) -- (c)
    (a) -- (d)
    (a) -- (g)
    (d) -- (e)
    (d) -- (f)
    (d) -- (g)
    (g) -- (h)
    (g) -- (i);
\end{tikzpicture}
\end{subfigure}
\hfill
%
% Middle graph
\begin{subfigure}{0.3\linewidth}
\centering
\begin{tikzpicture}[
  graph,
  baseline=(current bounding box.center)
]
  \node[vertex]           (top)    at ( 0, 1) {};
  \node[vertex]           (bottom) at ( 0,-1) {};
  \node[vertex, fill=red] (left)   at (-1, 0) {};
  \node[vertex, fill=red] (right)  at ( 1, 0) {};

  \draw
    (top) -- (bottom)
    (left) -- (right)
    (top) -- (left)
    (top) -- (right)
    (bottom) -- (left)
    (bottom) -- (right);
\end{tikzpicture}
\end{subfigure}
\hfill
%
% Right graph
\begin{subfigure}{0.3\linewidth}
\centering
\begin{tikzpicture}[
  graph,
  baseline=(current bounding box.center)
]
  \node[vertex, fill=green] (top)    at ( 0, 1) {};
  \node[vertex, fill=green] (bottom) at ( 0,-1) {};
  \node[vertex]             (left)   at (-1, 0) {};
  \node[vertex]             (right)  at ( 1, 0) {};

  \draw
    (top) -- (bottom)
    (left) -- (right)
    (top) -- (left)
    (top) -- (right)
    (bottom) -- (left)
    (bottom) -- (right);
\end{tikzpicture}
\end{subfigure}

\caption{The coloring of the starlink network topology and experiment set. The red nodes of the red experiment must be mapped to the red nodes on the starlink, and similarly for the green nodes on the green experiment. In each experiment, the uncolored nodes can be mapped to any node on the topology, including the colored ones.}
\label{fig:starlink_coloring}

\end{figure*}

We demonstrate the ability to train a policy given the coloring constraint described in Section \ref{networkmodel}. Figure \ref{fig:starlink_coloring} shows the specific coloring of the starlink network in our evaluations. Results on the colored dumbbell and grid topologies are left to Appendix \ref{other_colored_topologies}. The experiment durations are left unchanged. With this coloring, the policy must learn to overcome resource contention in the high-degree region of the network. Placing the red experiment requires at least one virtual link that consumes links along a path that includes the green nodes, which themselves must host the green experiment. This setup mimics a real-world scenario where a specific kind of hardware that exists at topologically-distant locations in the network is requested by an experiment at the same time as more centrally-located hardware is requested by another experiment.

\section{Results}\label{results}
We are interested in the ability of a policy to operate in varying noise levels driven by $\gamma$, primarily because it describes the effect of link activation probability on the ability to place experiments efficiently. In all subsequent plots in this section, 25 values of $\gamma$ were chosen between the interval $[1.5, 5.8]$ to evaluate 100 independent episodes, from which the success rate and the mean step count are calculated.  Appendix~\ref{appx:curriculum} describes the selection of $\mathcal{P}$ for each topology by comparing the performance after each phase of curriculum training.
\begin{figure*}[t] 
    \centering
    \includegraphics[width=\textwidth]{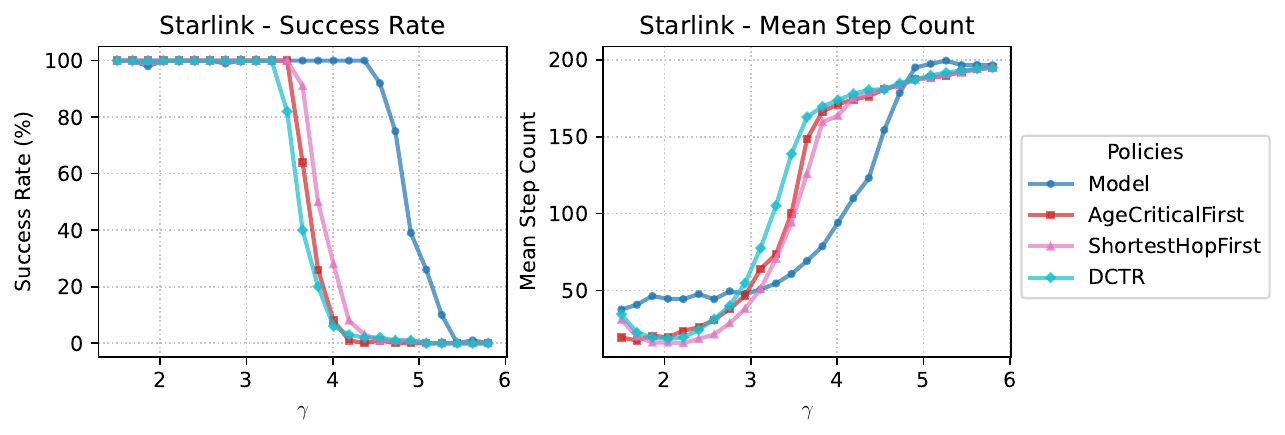} % Scale to full text width
    \caption{Comparison of success rate and mean step count between the starlink topology for $\mathcal{P}=9$ of the model, \textsc{AgeCriticalFirst}, \textsc{ShortestHopFirst}, and \textsc{DCTR}. }
    \label{fig:three_policies}
\end{figure*}

\subsection{Model vs Baselines on Uncolored Starlink Topology}
We compare $\mathcal{P}=9$ of the model to the baseline heuristics using the noise robustness test on the starlink topology.  Figure~\ref{fig:three_policies} shows that \textsc{AgeCriticalFirst}, \textsc{ShortestHopFirst}, and \textsc{DCTR} begin to fail at $\gamma \approx 3.5$ ($p_l \approx 0.030$), whereas the model begins to fail at $\gamma \approx 4.5$ ($p_l \approx 0.011$). At $\gamma = 4.0$ ($p_l \approx 0.018$), the model places all experiments in approximately 95 steps, whereas all heuristics require approximately 170 steps. Additionally, the model first fails at a 59\% lower probability than the best performing heuristic \textsc{AgeCriticalFirst}, and at a 66\% lower probability than the worst performing heuristic \textsc{DCTR}, indicating greater capability in high noise. For near-term quantum networks above $\gamma > 3.0$, the model succeeds faster, minimizing network resource consumption. The local strategy of \textsc{AgeCriticalFirst} and \textsc{ShortestHopFirst} greedily consumes active sublinks for virtual link generation as soon as they become available, inherently producing a greater proportion of unhelpful virtual links. Across most of the higher $\gamma$ range, representing near-term quantum network capabilities, the learned policy completes each episode in fewer steps on average and maintains a higher success rate.

\subsection{Model vs Baselines on Uncolored Dumbbell and Grid Topologies}
\begin{figure*}[t] 
    \centering
    \includegraphics[width=0.9\linewidth]{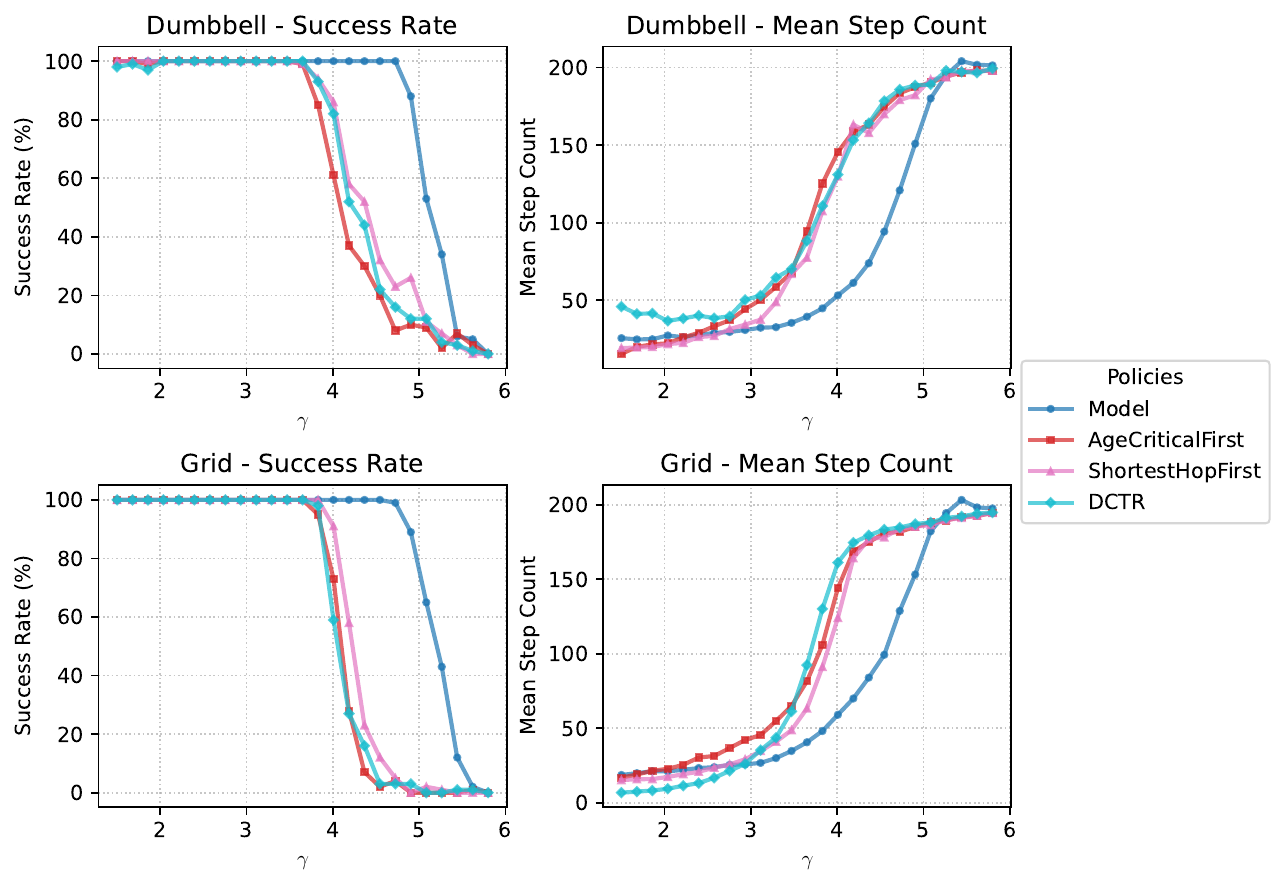}
    \caption{Success rate and mean step count as a function of $\gamma$, corresponding to link activation probability, on the dumbbell and grid topologies for $\mathcal{P}=8$ for both models, \textsc{AgeCriticalFirst}, \textsc{ShortestHopFirst}, and \textsc{DCTR}. }
    \label{fig:three_policies_combined}
\end{figure*}
Figure~\ref{fig:three_policies_combined} demonstrates the performance of the Deep Q-Network against baseline heuristics on the dumbbell and grid topologies. This highlights generalization across topologies featuring a bottleneck along the dumbbell's bridge node and path redundancy due to the high connectivity of the grid. On the dumbbell topology, the model first fails at 66\% lower probability than the best performing heuristic \textsc{DCTR}, and at 71\% lower probability than the worst performing heuristic \textsc{AgeCriticalFirst}.  The model also maintains a strictly lower mean step count across the operational regime. At $\gamma = 4.0$, the model places the experiments in approximately 50 steps, whereas the baseline heuristics require between 125 and 150 steps. In the grid topology, the model first fails at a 51\% lower probability than the best performing heuristic \textsc{ShortestHopFirst}, and at a 59\% lower probability than the worst performing heuristic \textsc{AgeCriticalFirst}. At $\gamma = 4.0$, the model takes around 55 steps to complete an episode, whereas the policies require the baselines require approximately between 125 and 160 steps.

\subsection{Model vs Baselines on Colored Starlink Topology}

\begin{figure*}[ht]
    \centering
    \includegraphics[width=\textwidth]{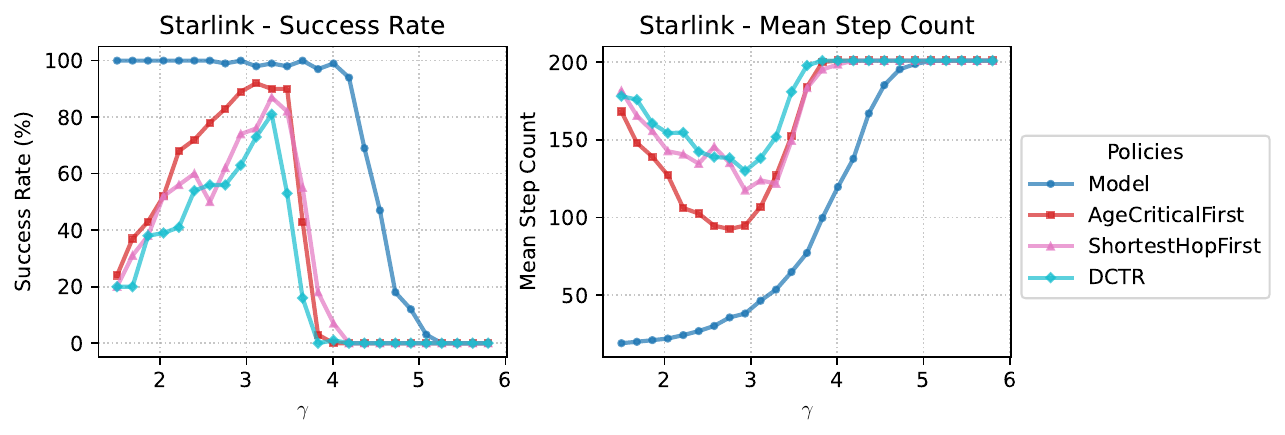}
    \caption{Comparison of success rate and mean step count between $\mathcal{P}=7$ of the model and baseline heuristics on the colored variant of the problem on the starlink topology. }
    \label{fig:starlink_colored_with_whites_three_policies}
\end{figure*}

We compare the model to the baseline algorithms on the colored starlink topology in Figure~\ref{fig:starlink_colored_with_whites_three_policies}. The figure shows the model succeeding frequently at low values of $\gamma$, where the heuristics cannot reliably succeed, likely because they exhaust the finite memories of the bottleneck node needed to form the required virtual link, whereas the model carefully generates its virtual links. In particular, the model begins to fail more than 20\% of episodes at a 51\% lower probability than the best performing heuristic \textsc{AgeCriticalFirst}, and at a 59\% lower probability than the worst performing heuristic \textsc{DCTR}. The model completes its episodes at least 60 steps quicker than the most efficient heuristic for all values of $\gamma < 4.1$.

\section{Behavior Profiling}\label{behavior_profiling}
\subsection{LLM Policy Distillation}\label{llm_policy_distillation}

\begin{figure*}[t]
\centering
\includegraphics[width=\textwidth]{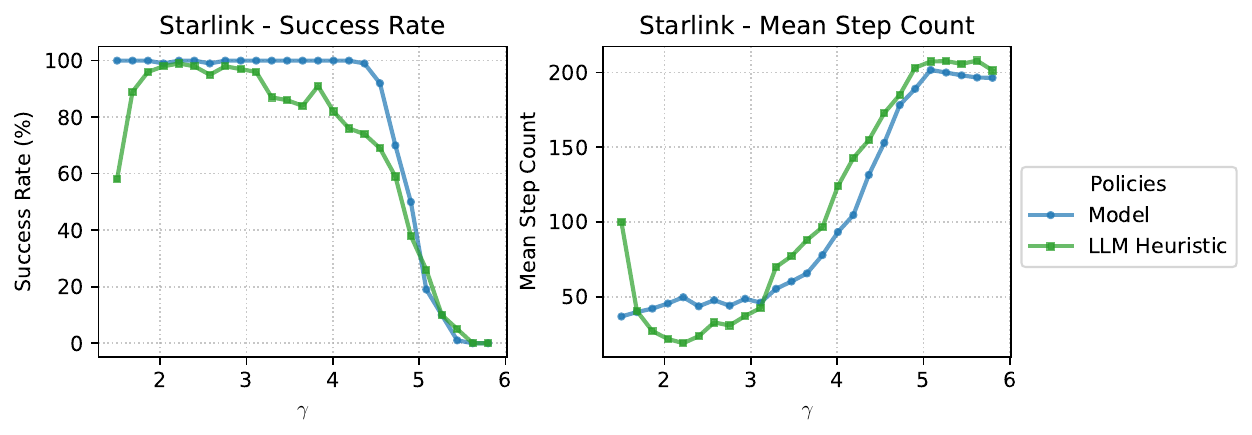}
\caption{Robustness to changes in link activation probability on the uncolored starlink topology for $\mathcal{P}=7$ of the model and the LLM-generated heuristic.}
\label{fig:starlink_vs_llm_heuristic}
\end{figure*}

We tasked Gemini 3.1 Pro with deriving a novel policy for the uncolored starlink topology ($N=9$), given examples of actions taken by the DQN-based policy. Details about the prompt are given in Appendix \ref{appendix:llm}. The policy returned by the LLM partitions the nodes into hub and leaf nodes, processes each experiment in order, identifies their optimal placements (prioritizing hub nodes), and places them if possible. If no experiment can be placed, then the virtual links useful for the optimal placements are generated. This process continues until all experiments of $\mathcal{X}$ have been placed. Figure \ref{fig:starlink_vs_llm_heuristic} shows the comparison of the LLM-based policy to the DQN-based policy. The DQN-based policy succeeds more frequently than the LLM-based one for $\gamma$ values less than $\gamma \approx 5.0$. For $\gamma > 3.2$, the DQN-based model completes the episodes in fewer steps than the LLM's heuristic. 
\subsection{Metrics}
We define three metrics to identify empirical differences between policies, enabling the identification of specific reasons why certain policies might succeed more frequently. \textsc{HoldingTime} measures the average time that a sublink is active and unused across all active links during an episode. \textsc{BridgeSpan} measures the average length of the shortest physical path between the two endpoints of all generated virtual links. \textsc{HubAnchorBias} measures the average maximum degree of the endpoints of a generated virtual link. Specifically, the average of $\max(\deg(u), \deg(v))$ is taken over the virtual links generated whose endpoints are $u, v \in N$, where the degree is taken with respect to the static physical edge set.
\begin{figure*}[b]
    \centering
    \includegraphics[width=\textwidth]{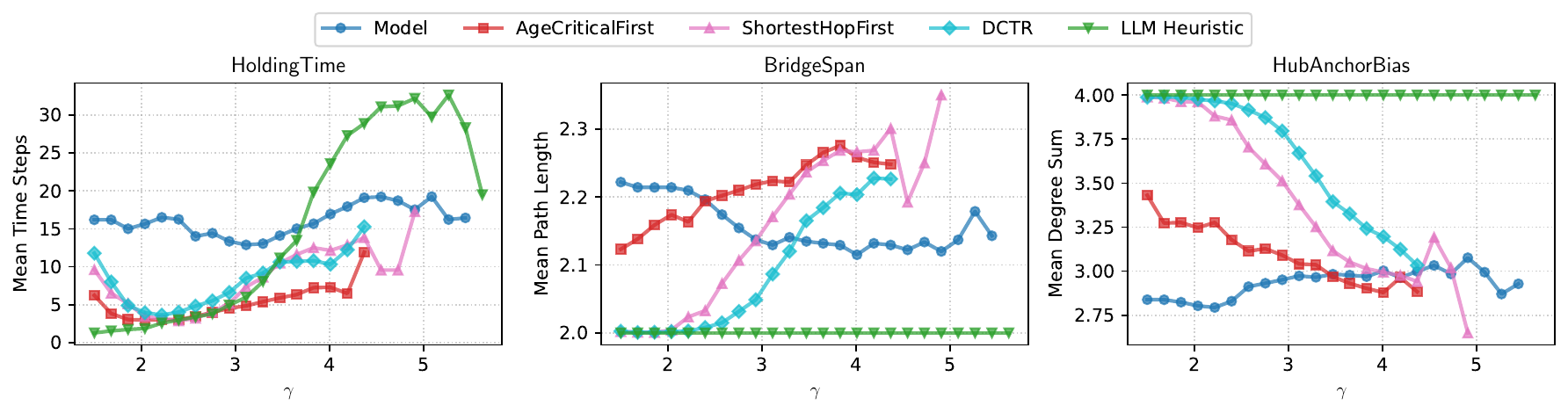} % Scale to full text width
    \caption{Three metrics \textsc{HoldingTime}, \textsc{BridgeSpan}, and \textsc{HubAnchorBias} measured on the uncolored starlink topology over successful episodes for $\mathcal{P}=7$ of the model, the baseline heuristics, and the LLM heuristic. Absent points are a result of a policy's 0\% success rate for that value of $\gamma$.}
    \label{fig:metrics_success_only_three_policies_behavioral_1x3}
\end{figure*}
Figure \ref{fig:metrics_success_only_three_policies_behavioral_1x3} compares these metrics on the starlink topology. The model learned a patient policy, as shown by its greater \textsc{HoldingTime} relative to the heuristics.  Even when physical links are activated more frequently (low $\gamma$), the model avoids greedily consuming virtual links and instead waits for more activated physical links before consuming them. Interestingly, the LLM waits longer than the model after $\gamma \approx 3.5$ while maintaining relatively high success rates. Next, in the operational regime, the model prefers to consistently generate virtual links of lesser span than those generated by the baseline heuristics. This result shows the model has more reliably learned to optimally place a $K_4$-based experiment. Note that in the starlink topology, the $K_4$ experiment requires at least two 2-path virtual links to be placed, and any longer links are unnecessary. The LLM clearly identified this requirement in designing its heuristic, as shown by its constant \textsc{BridgeSpan} of two. Lastly, the model demonstrates a low \textsc{HubAnchorBias} relative to all other policies, indicating that it prefers to schedule its experiments at peripheral nodes. On the other extreme, the LLM heuristic always generates virtual links with an endpoint in the hub, given its \textsc{HubAnchorBias} of four for all values of $\gamma$.

\section{Conclusion}

The work demonstrates an application of reinforcement learning techniques in probabilistic environments to quantum networks. Whereas prior results optimize connectivity-layer scheduling policies, our model optimizes link-layer scheduling tasks under realistic constraints, obtaining a policy that operates reliably even if link activation probabilities are low. Our results show that the use of reinforcement learning to obtain policies can ultimately aid the design and expansion of near-term quantum network topologies and the efficient management of end-user requests. We also give initial demonstrations of providing trajectories performed by a classically-trained model, whose policy itself is difficult to interpret, to a frontier LLM enabling it to create a novel policy that performs well and can described using natural language. Whereas directly measuring certain aspects of a policy's behavior  gives specific, targeted insights that can allude to a general policy, our work suggests that LLMs can be a useful tool for extracting a policy that is both easily described and useful in more general cases and in larger networks.

\section{AI Use Disclosure}
In this work, we used generative AI tools, in particular Gemini 3.1 Pro, to aid in implementing an initial version of the MPNN and the high-level training loop. We additionally conducted experiments with Gemini 3.1 Pro to obtain the results in Section \ref{llm_policy_distillation}. We have not used generative AI tools to interpret results or to design or provide feedback on research methodology or experiments, or to support qualitative or thematic data analysis. The uses of AI to generate synthetic data sets, to develop theoretical models or conceptual frameworks, to formulate mathematical claims, to assist with writing proofs, to propose or refine hypotheses, to assist with translation, and to clean or reformat the dataset, were not applicable to this work. Additionally, we used generative AI tools to write software code to generate the plots, and to create an initial version of Figure \ref{fig:virtual-link-placement}. We manually modified and extended the AI-assisted implementation to suit our problem formulation, simultaneously reviewing and revising the implementation. We take responsibility for the final content of this work, including text, claims or artifacts produced with the aid of generative AI.

\subsection*{Reproducibility statement}

To ensure the reproducibility of our training process, we provide the source code to train and evaluate our models at the following URL: \url{https://github.com/leonrode/simultaneous-entanglement-requests-in-quantum-networks}. The source code includes the environment and training configurations used for the results shown here. A README.md file is present in the repository with instructions.

\newpage
\bibliography{iclr2027_conference}
\newpage
\section{Appendix}
\appendix
\section{Double Deep Q-Learning}\label{appendix:DDQN}

Each training time step, the model either chooses a random action with probability $\epsilon$, or selects the action recommended by the online network. We use a dynamic epsilon-decay strategy, in which $\epsilon$ decays from $\epsilon_0$ to $\epsilon_n$ over an $n$-length schedule given by

\[
\epsilon_t = \epsilon_0 + \min\left(1, \frac{t}{n}\right)(\epsilon_n-\epsilon_0).
\]
The environment processes the action and the action, along with the reward received, is pushed to the replay buffer previously described. At a certain frequency, the online and target network are updated. A minibatch of 256 transitions is sampled from the replay buffer. Given a state transition $(s, a, r, s')$, the online network parameterized by $\boldsymbol{\theta}$ selects the best action $a^*$ for each next state $s'$ by taking the $\arg\max$ over all Q-values computed over $\mathcal{A}$:
$$
a^* = \arg\max_{a'} Q(s', a'; \boldsymbol{\theta})
$$
Then, the target network parameterized by $\boldsymbol{\theta}^-$ computes its Q-value for $a^*$ by $Q(s', a^*; \boldsymbol{\theta}^-)$ and the temporal difference is computed to give a ground-truth indicator of the value of selecting action $a$ in state $s$, given the true reward, $r$ received by taking action $a$.

\[
Y_t^{\text{DoubleQ}} = r + \gamma_{\text{RL}} \cdot Q\left(s', a^*; \boldsymbol{\theta}^-\right) (1 - \text{done})
\]
where $\text{done} \in \{0, 1\}$ is the terminal indicator, and $\gamma_{\text{RL}}$ is the discount factor. Then the online network computes $Q(s, a; \boldsymbol{\theta})$, its \textit{current} estimate for the value of that action, and $\boldsymbol{\theta}$ is updated by minimizing the Mean Squared Error loss against $Y_t^{\text{DoubleQ}}$ via Adam. Simultaneously, the target network parameters $\boldsymbol{\theta}^-$ undergo Polyak soft-updates at every training step according to:
\[
\boldsymbol{\theta}^- \leftarrow \tau \boldsymbol{\theta} + (1 - \tau) \boldsymbol{\theta}^-
\]
where $\tau \ll 1$ ensures that target Q-values transition smoothly without introducing moving-target instabilities.

\section{Observation Space}

The physical network state is encoded into a graph-structured observation dictionary to interface with the MPNN in Appendix \ref{appendix:MPNN}. All scalar features are normalized to $[0, 1]$ to stabilize gradient propagation. The observation tensor at time $t$ consists of:
\begin{itemize}
\item Node Features ($\mathbf{h}_v^t \in \mathbb{R}^1$): The number of available memories of node $v$, normalized by the maximum number of memories in the network.
\item Edge Features ($\mathbf{e}_{vw} \in \mathbb{R}^2$): The first dimension is the sublink's age scaled by the decoherence limit ($A_t(l) / m^\star$). The second dimension is the remaining execution time in the case that the link is locked by a placed experiment, normalized by $m^\star$.
\item Global Status ($\mathbf{g} \in \{0, 1\}^{\vert{}\mathcal{X}\vert{}}$): A binary indicator vector where $g_k = 1$ if experiment $X_k$ has not yet been placed, and $0$ if placed.

\end{itemize}

\section{Message Passing Neural Network}\label{appendix:MPNN}
Since the model learns from graph-structured information, we employ a Message Passing Neural Network (MPNN) defined in \citet{mpnn} for topology-aware decision-making. We encourage the reader to reference the work for conceptual details of MPNNs. The work defines a message function $M_t$ and a vertex update function $U_t$ used during the message-passing phase of the forward pass. Following their notation, our message function $M_t(\mathbf{h}^t_v, \mathbf{h}^t_w, \mathbf{e}_{vw})$ is parameterized by a 2-layer MLP with ReLU activations:
\[
M_t(\mathbf{h}^t_v, \mathbf{h}^t_w, \mathbf{e}_{vw}) = \text{MLP}_m(\mathbf{h}^t_v, \mathbf{h}^t_w, \mathbf{e}_{vw})
\]
where

\begin{align*}
\text{MLP}_m(\mathbf{h}^t_v, \mathbf{h}^t_w, \mathbf{e}_{vw}) = \, & \mathbf{W}_2^{(m)}  \text{ReLU}\Big(\mathbf{W}_1^{(m)} [\mathbf{h}_w^t \Vert \mathbf{h}_v^t \Vert \mathbf{e}_{vw}]
 + \mathbf{b}_1^{(m)}\Big) + \mathbf{b}_2^{(m)}
\end{align*}

Here, $\Vert$ denotes vector concatenation, $\mathbf{W}_1^{(m)} \in \mathbb{R}^{d \times (2d + d_e)}$ and $\mathbf{W}_2^{(m)} \in \mathbb{R}^{d \times d}$ are the weight matrices specific to the message-passing phase, and $\mathbf{b}_1^{(m)} \in \mathbb{R}^d$ and $\mathbf{b}_2^{(m)} \in \mathbb{R}^d$ are bias vectors. The term $d$ denotes the hidden representation dimension, and $d_e$ is the dimension of the edge features. In our setup, $d=128$ and $d_e = 2$. The messages directed at a given node $v$ are sum-pooled over its neighborhood $\mathcal{N}(v)$:
\[
\mathbf{m}^{(t+1)}_v = \sum_{w \in \mathcal{N}(v)} M_t(\mathbf{h}^t_v, \mathbf{h}^t_w, \mathbf{e}_{vw})
\]

Whereas $M_t$ is responsible for defining the communication transferred between adjacent nodes, the update function $U_t$ is responsible for updating each node's state as a consequence of the aggregated messages. $U_t$ acts on both the node's original hidden state $\mathbf{h}^t_v$ and the accumulated message $\mathbf{m}^{t+1}_v$ via a secondary MLP:
\[
\mathbf{h}_v^{t+1} = U_t(\mathbf{h}_v^t, \mathbf{m}_v^{(t+1)}) = \text{MLP}_u([\mathbf{h}_v^t \Vert \mathbf{m}_v^{(t+1)}])
\]

In our configuration, message-passing and updating occurs for $T=3$ layers.

The final phase of the forward pass involves a readout function $R$, which computes a global graph embedding to serve as the input to the discrete action head. The readout function produces the vector $\hat{\mathbf{y}}_t$ containing the Q-values for each action. We utilize a permutation-invariant mean-pooling operation over the final node states, which is then concatenated with a global status vector $\mathbf{g}$: $\hat{\mathbf{y}}_t = R(\{\mathbf{h}^{(T)}_v \mid v \in N\}, \mathbf{g}) = \text{MLP}_q(\mathbf{x}_\text{readout})$ 
where the fused readout representation is defined as $\mathbf{x}_\text{readout} = \left[ \frac{1}{|N|} \sum_{v \in N} \mathbf{h}_v^{(T)} \;\Vert\; \mathbf{g} \right]$. The Q-value projection head evaluates this representation using

{\footnotesize
\[
\text{MLP}_q(\mathbf{x}_\text{readout}) = \mathbf{W}_2^{(q)} \text{ReLU}\Big(\mathbf{W}_1^{(q)} \mathbf{x}_\text{readout} + \mathbf{b}_1^{(q)}\Big) + \mathbf{b}_2^{(q)}
\]} 

where $\mathbf{W}_1^{(q)} \in \mathbb{R}^{d \times (d + |\mathcal{X}|)}$, $\mathbf{W}_2^{(q)} \in \mathbb{R}^{|\mathcal{A}| \times d}$, $\mathbf{b}_1^{(q)} \in \mathbb{R}^d$, and $\mathbf{b}_2^{(q)} \in \mathbb{R}^{|\mathcal{A}|}$. Here, $|\mathcal{A}|$ is the total size of the discrete action space, and $|\mathcal{X}|$ is the number of scheduled experiments tracked by $\mathbf{g}$. Before decision-making, $\hat{\mathbf{y}}_t$ is action-masked to heavily penalize illegal actions. Illegal actions have their Q-values modified to $-10^9$ in the action head before action selection.

\section{Performance on Colored Dumbbell and Grid Topologies}\label{other_colored_topologies}

We test the training procedure on colored variants of the dumbbell and grid topologies, whose colorings are shown below:

\begin{figure*}[h!]
    \begin{subfigure}{0.45\linewidth}
\centering
\begin{tikzpicture}[
  graph,
  baseline=(current bounding box.center)
]
  % Central path
  \node[vertex, fill=lightgray]             (center) at ( 0, 0) {};
  \node[vertex, fill=green] (left)   at (-1, 0) {};
  \node[vertex, fill=green] (right)  at ( 1, 0) {};

  % Left diamond
  \node[vertex, fill=lightgray]           (lefttop)    at (-2, 1) {};
  \node[vertex, fill=red] (leftmiddle) at (-3, 0) {};
  \node[vertex, fill=lightgray]           (leftbottom) at (-2,-1) {};

  % Right diamond
  \node[vertex, fill=lightgray]           (righttop)    at (2, 1) {};
  \node[vertex, fill=red] (rightmiddle) at (3, 0) {};
  \node[vertex, fill=lightgray]           (rightbottom) at (2,-1) {};

  \draw
    (leftmiddle) -- (lefttop)
    (leftmiddle) -- (leftbottom)
    (lefttop) -- (left)
    (leftbottom) -- (left)
    (left) -- (center)
    (center) -- (right)
    (right) -- (righttop)
    (right) -- (rightbottom)
    (righttop) -- (rightmiddle)
    (rightbottom) -- (rightmiddle);
\end{tikzpicture}
\end{subfigure}
\hfill
%
% Right graph
\begin{subfigure}{0.45\linewidth}
\centering
\begin{tikzpicture}[
  graph,
  baseline=(current bounding box.center)
]
  % Top row
  \node[vertex, fill=red]       (n02) at (-1, 2) {};
  \node[vertex, fill=lightgray] (n12) at ( 0, 2) {};
  \node[vertex, fill=green]     (n22) at ( 1, 2) {};

  % Middle row
  \node[vertex, fill=lightgray] (n01) at (-1, 1) {};
  \node[vertex, fill=lightgray] (n11) at ( 0, 1) {};
  \node[vertex, fill=lightgray] (n21) at ( 1, 1) {};

  % Bottom row
  \node[vertex, fill=green]     (n00) at (-1, 0) {};
  \node[vertex, fill=lightgray] (n10) at ( 0, 0) {};
  \node[vertex, fill=red]       (n20) at ( 1, 0) {};

  \draw
    % Horizontal edges
    (n02) -- (n12) -- (n22)
    (n01) -- (n11) -- (n21)
    (n00) -- (n10) -- (n20)

    % Vertical edges
    (n02) -- (n01) -- (n00)
    (n12) -- (n11) -- (n10)
    (n22) -- (n21) -- (n20);
\end{tikzpicture}
\end{subfigure}
    \centering

\end{figure*}

\begin{figure*}
    \includegraphics[width=\textwidth]{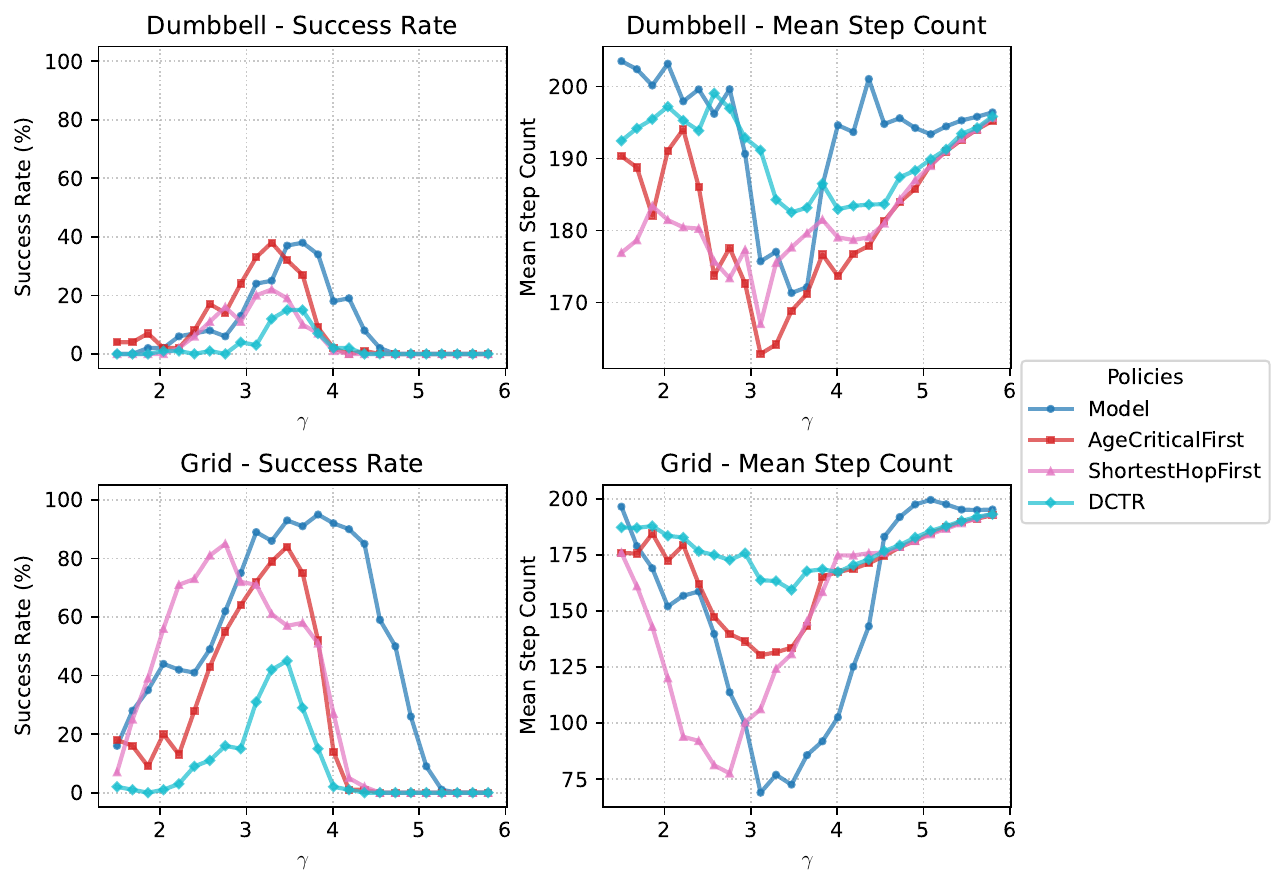}
    \caption{Comparison of the robustness to changes in link activation probability on colored dumbbell and grid topologies, with $\mathcal{P}=6$ and $\mathcal{P} = 7$ of the model, respectively.}
    \label{fig:betweenness_reward_dumbbell_colored_grid_colored_gamma_sweep_three_policies}

\end{figure*}

The experiment colorings are the same as in the colored starlink case. Here, to place the red and green $K_4$ experiments, the agent must generate virtual links between the red and green nodes at opposite corners. Thus the required physical links along the intermediate paths are contested, and the agent needs to learn when to optimally generate virtual links accordingly. Figure \ref{fig:betweenness_reward_dumbbell_colored_grid_colored_gamma_sweep_three_policies} shows the performance on the colored variant of the dumbbell and grid topologies. 

In the dumbbell case, there are limited routes to generate the virtual links needed to support a $K_4$ topology. The model performs similarly to the \textsc{AgeCriticalFirst} heuristic, which most directly relieves the  $m^\star$ constraint. In the region where the model outperforms \textsc{AgeCriticalFirst}, we see the model's greater episode lengths. The model performs with at least a 20\% success rate for 16\% lower link activation probability than the best-performing heuristic \textsc{AgeCriticalFirst}, and a 90\% lower link activation probability than the worst-performing heuristic, \textsc{DCTR}, which does not reach above a 20\% success rate for any $\gamma$. 

In the grid case, the model maintains above an 80\% success rate for 59\% lower link activation probabilities than the best-performing heuristic \textsc{AgeCriticalFirst}, and 94\% lower link activation probabilities than the worst-performing heuristic \textsc{DCTR}, which does not reach above 80\% success.

\section{Curriculum Phase Evolution}\label{appx:curriculum}

\subsection{Uncolored Variant}
\begin{figure*}
    \includegraphics[width=\textwidth]{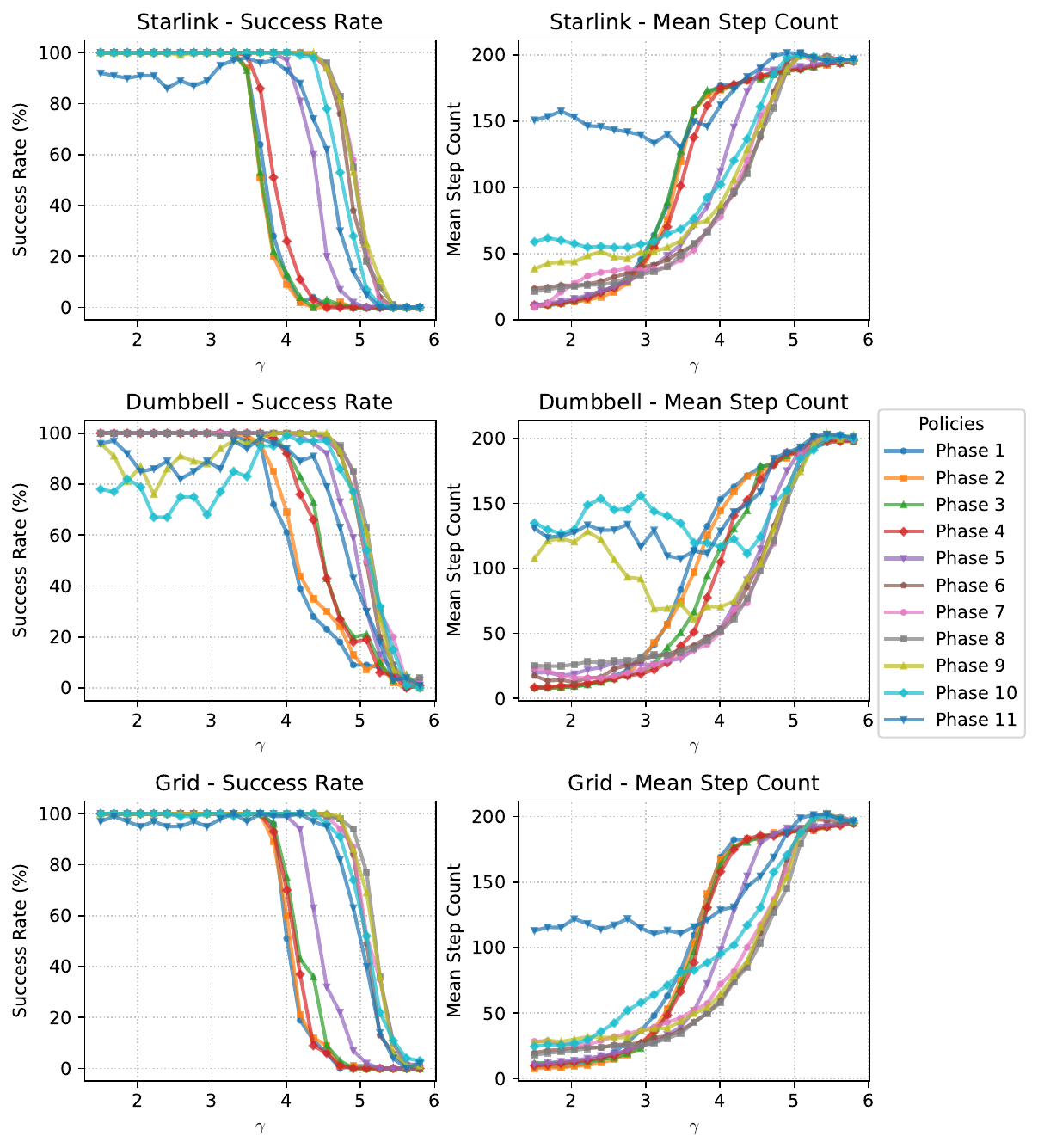}
    \caption{Comparison of the robustness to changes in link activation probability for each phase of the curriculum training process for each uncolored topology. We find that ending the curriculum training early at $\mathcal{P}=9$, $\mathcal{P}=8$, and $\mathcal{P}=8$ for the starlink, dumbbell, and grid topologies respectively provides the most robustness to changes in noise. The training experience in higher noise environments appear to degrade the policy's ability to perform in low-noise settings.}
    \label{fig:betweenness_reward_starlink_uncolored_dumbbell_uncolored_grid_uncolored_gamma_sweep}

\end{figure*}

Figure \ref{fig:betweenness_reward_starlink_uncolored_dumbbell_uncolored_grid_uncolored_gamma_sweep}, shows a comparison of robustness to noise between all 11 phases of the curriculum training on the uncolored variant on the starlink, dumbbell, and grid topologies. Recall that phase $\mathcal{P}>1$ is used to seed the experience replay buffer for phase $\mathcal{P} + 1$, i.e. phase $\mathcal{P}$ bootstraps phase $\mathcal{P}+1$. The figures clarify whether the model retains its ability to succeed in low-noise settings even when subjected to high-noise in later phases. Whereas phases $\mathcal{P}\leq9$ succeed perfectly in low $\gamma$ settings in the starlink topology, later phases begin to deteriorate in performance, indicating that experiences in high noise override the policy learned in earlier phases.

\subsection{Colored Variant}\label{appendix:starlink_colored_phase_evolution}

\begin{figure*}
    \includegraphics[width=\textwidth]{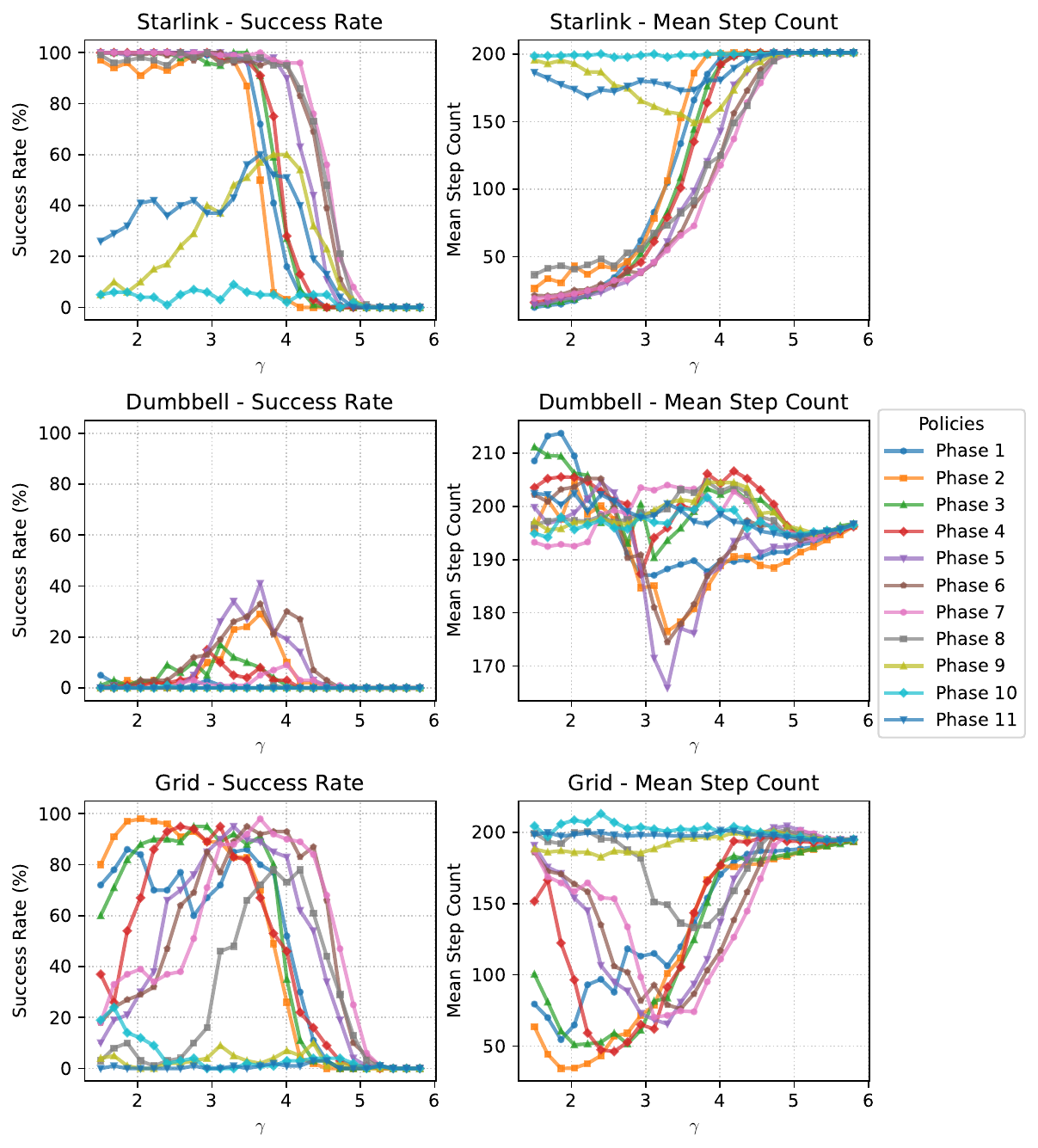}
    \caption{Comparison of the robustness to changes in link activation probability for each phase of the curriculum training process for the colored problem variant on each topology. We find that ending the curriculum training early at $\mathcal{P}=7$, $\mathcal{P}=6$, and $\mathcal{P}=7$ respectively provided the greatest robustness to changes in noise on the starlink, dumbbell, and grid topologies. }
    \label{fig:betweenness_reward_starlink_colored_dumbbell_colored_grid_colored_gamma_sweep}

\end{figure*}

Figures\ref{fig:betweenness_reward_starlink_colored_dumbbell_colored_grid_colored_gamma_sweep} shows a comparison of the robustness to noise between each of 11 phases of curriculum training on the colored variant on each topology. Contrary to the uncolored case, early phases were more robust than later phases. 

\section{LLM Prompt}\label{appendix:llm}

To generate the LLM heuristic for the uncolored starlink environment, we supplied a version of Gemini 3.1 Pro available in September 2026 the following prompt:

\begin{lstlisting}
You are an expert policy extractor whose task is to analyze the given execution traces of an agent acting on a starlink topology with N=9 nodes attempting to minimize the steps needed to assign a set of experiments. Derive a general heuristic that will outperform the model on a starlink constructed using 3 hub nodes but with any number of leaf nodes evenly distributed among the hub nodes.

The adjacency matrix for the host network of the execution traces is:
[[0, 1, 1, 0, 0, 1, 0, 0, 1],
[1, 0, 1, 0, 1, 0, 0, 1, 0],
[1, 1, 0, 1, 0, 0, 1, 0, 0],
[0, 0, 1, 0, 0, 0, 0, 0, 0],
[0, 1, 0, 0, 0, 0, 0, 0, 0],
[1, 0, 0, 0, 0, 0, 0, 0, 0],
[0, 0, 1, 0, 0, 0, 0, 0, 0],
[0, 1, 0, 0, 0, 0, 0, 0, 0],
[1, 0, 0, 0, 0, 0, 0, 0, 0]]

The set of experiments to be assigned formatted as (adjacency matrix, duration) was:
{([[0, 1, 1, 1], [1, 0, 1, 1], [1, 1, 0, 1], [1, 1, 1, 0]], 1), ([[0, 1, 1, 1], [1, 0, 1, 1], [1, 1, 0, 1], [1, 1, 1, 0]], 1)}

The maximum age of any link was m* = 52

The probability of physical link activation was p_link = 0.0111

5 parallel physical links exist between nodes, each of which is activated at each time step with probability p_link.

Virtual links are deterministically generated by consuming physical links along the shortest paths between nodes, and exist for m* before deactivating.

Output your heuristic as pseudocode.
\end{lstlisting}

The following is an excerpt from the execution traces provided to the LLM, specifically the first episode. The LLM can infer the time between virtual link generations, correspond their endpoint indices to the host network based on the adjacency matrix provided, and view the consequence of each action in terms of reward and episode result.

\begin{lstlisting}
    === EPISODE 1/15 | GAMMA: 4.5 | SEED: 42 ===
  Step 000 | Action: WAIT                           | Reward: -15.000
  Step 001 | Action: WAIT                           | Reward: -15.000
  ...
  Step 008 | Action: WAIT                           | Reward: -15.000
  Step 009 | Action: WAIT                           | Reward: -15.000
  Step 010 | Action: WAIT                           | Reward: -15.000
  Step 011 | Action: GENERATE_VL(0, 6)              | Reward: -15.000
  Step 012 | Action: WAIT                           | Reward: -15.000
  ...
  Step 021 | Action: WAIT                           | Reward: -15.000
  Step 022 | Action: WAIT                           | Reward: -15.000
  Step 023 | Action: WAIT                           | Reward: -15.000
  Step 024 | Action: WAIT                           | Reward: -15.000
  Step 025 | Action: GENERATE_VL(1, 8)              | Reward: -20.000
  Step 026 | Action: WAIT                           | Reward: -15.000
  Step 027 | Action: WAIT                           | Reward: -15.000
  Step 028 | Action: WAIT                           | Reward: -15.000
  Step 029 | Action: WAIT                           | Reward: -15.000
  Step 030 | Action: WAIT                           | Reward: -15.000
  ...
  
  Step 109 | Action: WAIT                           | Reward: -15.000
  Step 110 | Action: WAIT                           | Reward: -15.000
  Step 111 | Action: WAIT                           | Reward: -15.000
  Step 112 | Action: PLACE_EXPERIMENT(idx=0, offset=11) | Reward: +385.000
  Step 113 | Action: WAIT                           | Reward: -15.000
  ...
  Step 193 | Action: WAIT                           | Reward: -15.000
  Step 194 | Action: WAIT                           | Reward: -15.000
  Step 195 | Action: PLACE_EXPERIMENT(idx=1, offset=32) | Reward: +385.000
  Result: TERMINATED (Success) | Total Reward: -2274.000 | Total Steps: 196
\end{lstlisting}

\begin{figure*}
    \includegraphics[width=\textwidth]{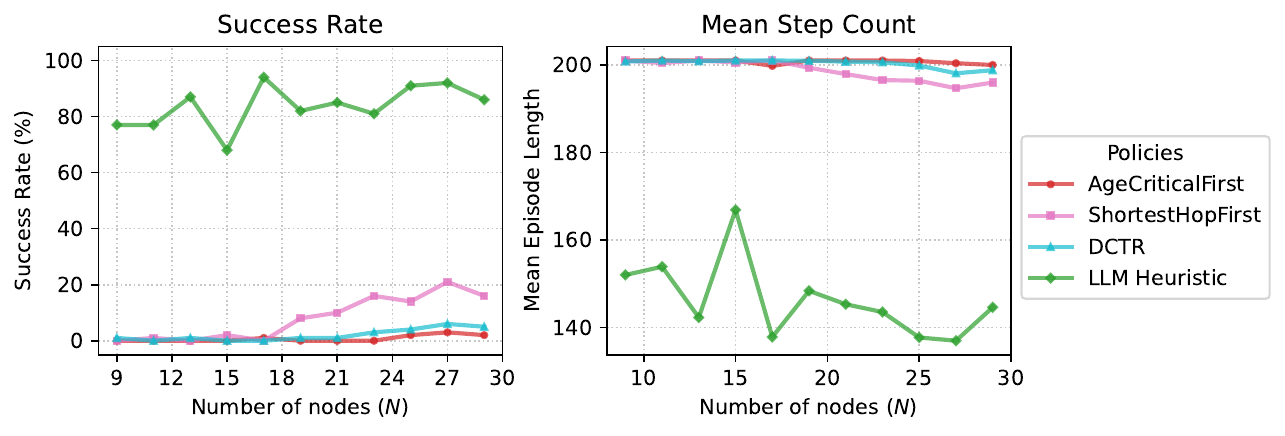}
    \caption{Comparison of the success rate and mean episode length for each heuristic as a function of topology size on the uncolored starlink topology. Each episode was run at constant $\gamma = 4.5$ ($p_l \approx 0.011$)}
    \label{fig:betweenness_reward_starlink_uncolored_n_sweep_llm_three_policies.pdf}

\end{figure*}

We also compared the success rates of the LLM heuristic against the baseline heuristic for increasing sizes of the uncolored starlink topology in Figure \ref{fig:betweenness_reward_starlink_uncolored_n_sweep_llm_three_policies.pdf}. In these tests, the number of central hub nodes remained fixed at three, so that the additional nodes were evenly distributed on the peripheral nodes. We note that the LLM generated heuristic performs well independently of network size, indicating that once a heuristic is obtained from the LLM, it can be reused as the network expands to include more nodes.

\end{document}